\documentclass[sn-mathphys-num]{sn-jnl}

\usepackage{graphicx}%
\usepackage{multirow}%
\usepackage{amsmath,amssymb,amsfonts}%
\usepackage{amsthm}%
\usepackage{mathrsfs}%
\usepackage[title]{appendix}%
\usepackage{xcolor}%
\usepackage{textcomp}%
\usepackage{manyfoot}%
\usepackage{booktabs}%
\usepackage{algorithm}%
\usepackage{algorithmicx}%
\usepackage{algpseudocode}%
\usepackage{listings}%

\usepackage{amsmath}
\usepackage{amssymb}
\usepackage{graphicx} 

\usepackage{dsfont}
\newcommand{\openone}{\mathds{1}}

\newcommand{\ket}[1]{\vert #1 \rangle}

\newcommand{\ketbra}[2]{\vert #1 \rangle \langle #2 \vert}
\newcommand{\abs}[1]{\vert #1 \vert}
\newcommand{\tr}{\operatorname{tr}}

\renewcommand{\Pr}{\mathrm{Pr}}

\renewcommand{\d}{\operatorname{d}}

\usepackage{xcolor}

\newcommand{\new}[1]{{ #1}}

\newcommand{\hide}[1]{{\color{blue}}}

\theoremstyle{thmstyleone}%
\theoremstyle{thmstyletwo}%

\theoremstyle{thmstylethree}%

\begin{document}

\title{\mbox{Ultradecoherence model of the measurement process}}


\author*{\fnm{Hai-Chau} \sur{Nguyen}}\email{chau.nguyen@uni-siegen.de}

\affil*{\orgdiv{Naturwissenschaftlich--Technische Fakult\"{a}t}, \orgname{Universit\"{a}t Siegen}, \orgaddress{\street{Walter-Flex-Stra{\ss}e 3}, \city{Siegen}, \postcode{57068}, \state{NRW}, \country{Germany}}}


\abstract{
It is proposed that measurement devices can be modelled to have an open decoherence dynamics that is faster than any other relevant timescale, which is referred to as the ultradecoherence limit.
\new{In this limit, the measurement device always assumes a definite state upto the accuracy set by the fast decoherence timescale.} 
\new{Further,} it is shown that the clicking rate of measurement devices can be derived from its underlying parameters, not only for the von Neumann ideal measurement devices but also for photon detectors in equal footing.
This study offers a glimpse into the intriguing physics of measurement processes in quantum mechanics, with many aspects open for further investigation.}

\keywords{quantum mechanics, measurement, decoherence, time, von Neumann}

\maketitle

\section{Introduction}

\new{Since the formulation of quantum mechanics~\cite{born_zur_1926b,schrodinger_quantisierung_1926,born_zur_1926}, measurements have been always at the centre of the debate on the foundations of the theory. 
Indeed, while the dynamical evolution of a quantum mechanical system is unitary as dictated by the Schr\"odinger equation~\cite{schrodinger_quantisierung_1926}, measurements in quantum theory seem to follow rather different rules~\cite{born_zur_1926}.

The formal procedure of a measurement in quantum theory can be formulated as follow. 
Abstractly, a measurement apparatus is a device with a pointer $\mu$, which can point to values $\mu=0,1,\ldots,M$.
To start a measurement, the device is first set to be `ready,' where its pointer points to $\mu=0$. The quantum system to be measured, assumed to be in quantum state $\ket{\psi}$, is then coupled to the device. As a result of this coupling, the pointer of the device `clicks' on one of `outcomes' $\mu=1,2,\ldots,M$ at random. The probability for a particular outcome to happen is determined by Born's rule for measurement statistics~\cite{born_zur_1926}. The Born rule for measurement statistics dictates that the outcomes of the device $\mu=1,2,\ldots,M$ are associated with states $\{\ket{s_\mu}\}_{\mu =1}^{M}$ of the measured system, which form an orthonormal basis for the expansion $\ket{\psi} = \sum_{\mu} c_\mu \ket{s_\mu}$.~\footnote{Generalised measurements described by postive operator measures can be treated as this ideal measurement made on an appropriate ancillary system.} The probability for the outcome $\mu$ is then given by $\abs{c_\mu}^2$. In certain cases, the Born rule for measurement statistics can be accompanied by Born's rule for post-measurement states, which postulates that the measured system assumes state $\ket{s_\mu}$ once the outcome $\mu$ is observed.}

\new{Since the early development of quantum mechanics, there were already attempts to reconcile such a measurement process with the unitary dynamics described by the Schr\"odinger equation of quantum theory.}
A unitary model for the measurement process is dated back to von Neumann~\cite{von_neumann_mathematical_2018}. 
He regarded the ready state and the measurement outcomes of a measurement device as its different quantum states, \new{denoted by $\ket{0}$ and $\{\ket{\mu}\}_{\mu=1}^{M}$, respectively}. 
It \new{was} then argued that, for the measurement device to function as expected, the joint unitary evolution of the device plus the measured system must be such that states $\ket{0} \otimes \ket{s_{\mu}}$ evolve into $\ket{\mu} \otimes \ket{s_{\mu}}$~\cite{von_neumann_mathematical_2018}. \new{As a consequence of the linearity of the unitary evolution, if the system is initially in state $\ket{\psi} = \sum_{\mu} c_{\mu} \ket{s_\mu}$, the measurement device and the system then evolve as} 
\begin{equation}
\ket{0} \otimes \sum_{\mu} c_{\mu} \ket{s_\mu} \to 
\sum_{\mu} c_{\mu} \ket{\mu} \otimes \ket{s_\mu}.
\label{eq:von-neumann-sum}
\end{equation}
In the modern language, the device and the system get into an entangled state, referred to as a \emph{premeasurement state}~\cite{schlosshauer_decoherence_2005,zurek_decoherence_2003}.  
Unfortunately, eq.~\eqref{eq:von-neumann-sum} \new{raises various issues for the foundations} of quantum mechanics; see Ref. ~\cite{schrodinger_gegenwartige_1935,everett_relative_1957,wigner_remarks_1995,deutsch_quantum_1985,schlosshauer_decoherence_2005,zurek_decoherence_2003,zeh_interpretation_1970,zeh_toward_1973,zurek_pointer_1981,zurek_environment-induced_1982,presilla_measurement_1996,schlosshauer_decoherence_2007,kiefer_quantum_2022,schlosshauer_elegance_2011} and the references therein.

The premeasurement state~\eqref{eq:von-neumann-sum} results from the assumption that the measured system and the measurement device are isolated and thus follow a unitary evolution~\cite{von_neumann_mathematical_2018}. 
Nowadays, it has become clear that the environment surrounding the measurement device causes a fast decoherence and the dynamics of the measurement device is generally not unitary~\cite{zurek_decoherence_2003,schlosshauer_decoherence_2005,schlosshauer_decoherence_2007,kiefer_quantum_2022}. 
Yet, the premeasurement state~\eqref{eq:von-neumann-sum} is still widely assumed in the standard literature~\cite{kiefer_quantum_2022,von_neumann_mathematical_2018,schlosshauer_decoherence_2005,schlosshauer_decoherence_2007,schlosshauer_elegance_2011,zurek_decoherence_2003,schrodinger_gegenwartige_1935,everett_relative_1957,deutsch_quantum_1985,wigner_remarks_1995,zeh_interpretation_1970,zeh_toward_1973,zurek_pointer_1981,zurek_environment-induced_1982}. This means, one is implicitly assuming that the measured system and the measurement device can be considered to be isolated in a short time scale before the decoherence taking place.
 
It is, however, more natural to expect that most measurement devices are in a very strong \new{influence from} the environment at all time; so strong that quantum decoherence takes place much faster than any other relevant timescale. 
They will be referred to as being in `ultradecoherence.' 
Within the accuracy of its decoherence time, an ultradecohered system \new{is} always in a \emph{definite state}.
It is shown that, although the measured system does not get entangled with such an ultradecohered measurement device into the premeasurement state~\eqref{eq:von-neumann-sum}, it can trigger a transition of the device to a different state---a \emph{measurement click}. 
While the interaction with the measured system is of quantum mechanical nature, the resulted measurement click appears to behave much like a classical stochastic transition. 
The statistics of the measurement clicks follows the normal Born rule for measurement statistics~\cite{born_zur_1926}. 
Moreover, the transition of the state of the measurement device is accompanied by a transition in the state of the measured system, \new{encompassing} also the Born rule for post-measurement states~\cite{von_neumann_mathematical_2018}.

This simple model indicates that behind the familiar Born rules for quantum measurements~\cite{born_zur_1926,von_neumann_mathematical_2018} is the rich physics of the measurement process. 
In fact, for the ultradecohered measurement devices, the Born rules emerge dynamically. 
Here a measurement device can be characterised by a set of transition operators $\{R_{\mu 0}\}_{\mu=1}^{M}$. 
Upon interaction with the measured system, assumed to be in mixed state $\rho$, the measurement device can undergo a transition from the ready state $\ket{0}$ to one of the pointer states $\{\ket{\mu}\}_{\mu=1}^{M}$, stochastically in continuous time with rate $W_{\mu 0}$ given by
\begin{equation}
    W_{\mu 0} = \tr(\rho R_{\mu 0}^\dagger R_{\mu 0}).
    \label{eq:continuous-measurement-rate}
\end{equation}
After the transition, the state of the measured system assumes
\begin{equation}
    \sigma_{\mu} =  \frac{R_{\mu 0} \rho R_{\mu 0}^\dagger}{\tr(\rho R_{\mu 0}^\dagger R_{\mu 0})}.
    \label{eq:continuous-measurement-state}
\end{equation}
Here, one is only concerned with the first measurement click; after this first click, another conditional process is initiated to reset the device for a new measurement.

The dynamical rules~\eqref{eq:continuous-measurement-rate} and~\eqref{eq:continuous-measurement-state} are in fact known in practice to compute current fluctuations in quantum optics and quantum electronics~\cite{landi_current_2024}. There, the clicking rates are interpreted from the formal process of `unravelling' the open dynamics of the measured system, effectively assuming that the measurements are made on the `farfield emission'~\cite{landi_current_2024}. Here, we consider their emergence from the dynamics of the measurement devices themselves, which include the von Neumann ideal measurement devices as well as photon detectors.

\section{The ultradecoherence dynamics}

Consider two coupled quantum mechanical systems denoted by $\mathrm{D}$ (device) and $\mathrm{Q}$ (quantum) with
the total Hamiltonian 
\begin{equation}
    H= H_{\mathrm{D}} + H_{\mathrm{Q}}  + V,
\end{equation}
where $H_{\mathrm{D}}$ and $H_{\mathrm{Q}}$ are respectively the free Hamiltonians of system $\mathrm{D}$ and $\mathrm{Q}$, and $V$ their interaction term. 
The eigenstates of $H_{\mathrm{D}}$ and their energies are denoted by $\{\ket{\mu}\}_{\mu=0}^{M}$ and $\{\Omega_{\mu}\}_{\mu=0}^{M}$, respectively.
Generally, one can write
\begin{equation}
V =  \sum_{\mu \nu}  \ketbra{\mu}{\nu} \otimes V_{\mu \nu},
\end{equation} 
where the operator $V_{\mu \nu}$ intuitively describes the coupling of the transition of system $\mathrm{D}$ from state $\ket{\nu}$ to $\ket{\mu}$ to the dynamics of system $\mathrm{Q}$. 
As $V$ is hermitian, one has $V_{\mu \nu} = V^\dagger_{\nu \mu}$. It is also assumed $V_{\mu \mu} =0$ for simplicity. The natural quantum unit is assumed, so that $\hbar=1$ in all expressions.

Further, it is assumed that system $\mathrm{D}$ \new{constantly} couples to the environment at all time, which results in a fast dephasing dynamics in the basis $\{\ket{\mu}\}_{\mu=0}^{M}$.
Following the literature on open quantum systems~\cite{schlosshauer_decoherence_2007}, $\{\ket{\mu}\}_{\mu =0}^{M}$ is referred to as the \emph{preferred basis}.
While the microscopic mechanism is interesting in its own-right~\cite{schlosshauer_decoherence_2007,schlosshauer_quantum_2019,anglin_deconstructing_1997}, it suffices for our purpose to assume that the Lindblad master equation~\cite{schlosshauer_decoherence_2007} describing the open dynamics of the density operator $\rho^I$ of the joint system $\mathrm{DQ}$ in the interaction picture to be
\begin{equation}
    \frac{\mathrm{d} \rho^I}{\mathrm{d} t} = - i [V^I,\rho^I] + \sum_{\mu} \gamma_{\mu} \left( P_{\mu} \rho^I P_{\mu}^\dagger - \frac{1}{2} \left\{ P_{\mu}^\dagger P_{\mu}, \rho^I \right\} \right),
    \label{eq:Q-lindblad-interaction}
    \end{equation}
where
 $V^I(t) = e^{+i (H_{\mathrm{D}} + H_{\mathrm{Q}}) t}V e^{-i (H_{\mathrm{D}} + H_{\mathrm{Q}}) t}$,  
and  
$P_{\mu} =  \ketbra{\mu}{\mu} \otimes \openone_{\mathrm{Q}}$ are the Lindblad operators with the associated rates $\gamma_{\mu}$ responsible for the fast dephasing of system $\mathrm{D}$. Notice that the validity of the Lindblad equation is assumed globally for the joint system \new{$\mathrm{DQ}$} in interaction.  

Due to the fast dephasing in the preferred basis, one can assume that the whole system is `essentially classical.' 
In particular, the system can be `checked' at any time to reveal its definite state, provided that this interference takes place at a timescale much longer than the decoherence timescale.
This qualifies the \emph{ultradecoherence} limit, and the system is said to be \emph{ultradecohered}.

Employing the preferred basis, the density operator can be expanded as
$\rho^I = \sum_{\mu \nu}  \ketbra{\mu}{\nu} \otimes \rho_{\mu \nu}^I$,
where $\rho_{\mu \nu}^I$ are (block) operators acting only on system $\mathrm{Q}$.
Notice that the matrix elements of the density operator of the ultradecohered system $\mathrm{D}$ in the interaction picture are given by $\tr(\rho^I_{\mu \nu})$. In particular, $p_\mu= \tr(\rho^I_{\mu \mu})$ is the probability for the ultradecohered system to be found in state $\ket{\mu}$.

From the Lindblad equation~\eqref{eq:Q-lindblad-interaction}, one obtains an explicit evolution equation for $\rho^I_{\mu \nu}$ as 
\begin{equation}
   \frac{\mathrm{d} \rho_{\mu \nu}^I}{\mathrm{d} t} = -i \sum_{\lambda} (e^{i \Omega_{\mu \lambda} t} V_{\mu \lambda}^I \rho^I_{\lambda \nu} - \rho^I_{\mu \lambda}  V_{\lambda \nu}^I e^{i \Omega_{\lambda \nu} t} )- \gamma_{\mu \nu} \rho^I_{\mu \nu}.
   \label{eq:Q-master-full-interaction-element}
\end{equation}
where $V_{\mu \nu}^I (t) =   e^{+iH_Q t} V_{\mu \nu} e^{-iH_Qt}$, $\Omega_{\mu \nu} = \Omega_\mu - \Omega_\nu$, and $\gamma_{\mu \nu} = (\gamma_{\mu}+\gamma_{\nu})/2$ for $\mu \ne \nu$ and $0$ otherwise.
From eq.~\eqref{eq:Q-master-full-interaction-element}, one sees that $\rho^I_{\mu \nu}$  ($\mu \ne \nu$) is driven by a fast damping term $- \gamma_{\mu \nu} \rho^I_{\mu \nu}$. As a result, these coherence blocks remain small and are adiabatically determined by the diagonal blocks. 

\new{To carry out the adiabatic elimination, one rewrites the implicit solution for the coherence $\rho^I_{\mu \nu}$ ($\mu \ne \nu$) using the method of Green's function as
\begin{align}
    \rho^I_{\mu \nu} (t) = -i \int_{0}^{t} \mathrm{d} \tau e^{-\gamma_{\mu \nu} \tau} \sum_{\lambda} \left[ e^{i \Omega_{\mu \lambda} \Delta t} V_{\mu \lambda}^I (\Delta t) \rho^I_{\lambda \nu} (\Delta t)   - \rho^I_{\mu \lambda} (\Delta t) V_{\lambda \nu}^I (\Delta t) e^{i \Omega_{\lambda \nu} \Delta t} \right],
    \label{eq:Q-coherence-green-function}
\end{align}
where $\Delta t = t - \tau$.
Here, it is assumed that the coherence blocks $\rho^I_{\mu \nu}$ all together vanish at the initial time $t=0$.
As the exponential function $e^{-\gamma_{\mu \nu} \tau}$ decays fast, one can assume  the so-called Markov approximation: on the one hand, one approximates $\rho^I_{\lambda \nu} (t - \tau) \approx \rho^I_{\lambda \nu} (t)$ and $\rho^I_{\mu \lambda} (t-\tau) \approx \rho^I_{\mu \lambda} (t)$, and on the other hand, one extends the integral over $\tau$ to infinity~\cite{schlosshauer_decoherence_2007}. 
Further, in the summation over $\lambda$, one can retain only the dominating diagonal blocks.      
Within these approximations, one obtains

\begin{equation}
    \rho^I_{\mu \nu} \approx - i   (K^I_{\mu \nu} \rho^I_{\nu \nu} - \rho_{\mu \mu}^I K^I_{\mu \nu}) e^{i\Omega_{\mu \nu} t},
    \label{eq:Q-coherence-perturbative}
\end{equation}
where 
$K^I_{\mu \nu} = e^{+i H_Q t} K_{\mu \nu} e^{-i H_Q t}$
with
\begin{equation}
    K_{\mu \nu} = \int_{0}^{+\infty} \mathrm{d} \tau e^{-(\gamma_{\mu \nu} + i \Omega_{\mu \nu}) \tau}  V_{\mu \nu}^I (-\tau).
    \label{eq:K-element}
\end{equation}
}

By substitution of the solution~\eqref{eq:Q-coherence-perturbative} into eq.~\eqref{eq:Q-master-full-interaction-element} and concentrate on the case $\mu = \nu$, one arrives at the dynamical equation for the diagonal blocks only,
\begin{equation}
    \frac{\mathrm{d} \rho^I_{\mu \mu}}{ \mathrm{d}t} = - \sum_{\lambda \ne \mu} (V_{\mu \lambda}^I K_{\lambda \mu}^I \rho^I_{\mu \mu} +\rho^I_{\mu \mu} K^I_{\mu \lambda}  V^I_{\lambda \mu}) + \sum_{\lambda \ne \mu} (V^{I}_{\mu \lambda} \rho^I_{\lambda \lambda} K^I_{\lambda \mu} + K^I_{\mu \lambda} \rho^I_{\lambda \lambda} V^I_{\lambda \mu}),
    \label{eq:Q-master-classical}
\end{equation}
where the assumption $V_{\mu \mu} = 0$ has been used to explicitly impose $\lambda \ne \mu$ in the summation.

Notice that in order to be consistent with the Markov approximation, 
the solution to eq.~\eqref{eq:Q-master-classical} must be such that $\rho^{I}_{\mu \mu}$ vary in much a slower timescale than the decoherence timescale $\mathcal{O} (1/\gamma_{\mu \nu})$.
Further, it is not obvious from~\eqref{eq:Q-master-classical} that the positivity of the density operator is guaranteed. In all of the models considered below, however, one has 
$K^I_{\mu \nu} \approx V^I_{\mu \nu}/\gamma_{\mu \nu}$,
and the positiveness of the density operator follows. 
This is referred to as the \emph{resonant formula} for $K^I_{\mu \nu}$. 
\new{This approximation is most clearly appreciated in the concrete example of the photon detector below.
}

\section{Emergence of classical stochasticity}
Recall that $p_\mu= \tr (\rho^I_{\mu \mu})$ is the probability for the ultradecohered system $\mathrm{D}$ to be found in state $\ket{\mu}$. The rate equation for this classical probability can be directly obtained from eq.~\eqref{eq:Q-master-classical},
\begin{equation}
    \frac{\mathrm{d} p_\mu}{ \mathrm{d} t} = -  \tr [\Gamma_{\mu}^I \rho^I_{\mu \mu} +\rho^I_{\mu \mu} (\Gamma_{\mu}^I)^\dagger] + \sum_{\lambda \ne \mu}  \tr(F^I_{\mu \lambda} \rho^I_{\lambda \lambda}).
    \label{eq:master-resonant-probability}
\end{equation}
where we introduce $F^I_{\mu \lambda} =  K^{I}_{\lambda \mu}  V^I_{\mu \lambda} 
    +V_{\lambda \mu}^I K_{\mu \lambda}^I$ and $\Gamma^I_{\mu} = \sum_{\lambda \ne \mu} V^I_{\mu \lambda} K^I_{\lambda \mu}$.
As in the classical theory of stochastic processes, one can identify the `gain term,'
  $W_{\mu \lambda} = \tr(F^I_{\mu \lambda} \rho^I_{\lambda \lambda})$,
as the rate for the ultradecohered system $\mathrm{D}$ to make a transition from state $\ket{\lambda}$ to state $\ket{\mu}$. The `loss term' $\tr [\Gamma_{\mu}^I \varrho^I_{\mu \mu} +\varrho^I_{\mu \mu} (\Gamma_{\mu}^I)^\dagger]$ on the other hand describes the rate at which the system leaves state $\ket{\mu}$. 

The following problem is crucial to the further discussion: suppose the ultradecohered system $\mathrm{D}$ is initially at state $\ket{\mu}$, what is the distribution of the (random) time $T_\mu$ at which it makes the first transition? The time $T_\mu$ is called the \emph{persistent time} at state $\ket{\mu}$. Further, how is such a first step distributed among all other states $\ket{\nu}$? This is called the \emph{first step distribution}, denoted by $\pi^{(1)}_{\nu \mu}$.  
These are classic questions in the classical theory of stochastic processes~\cite{stirzaker_stochastic_2005}. 
\new{That the methods for them are applicable also to ultradecohered systems should be considered as an extrapolation, which, in principle, can only be confirmed by means of experiments.}

\new{Notice that the persistent probability $\Pr(T_\mu \ge t + \delta t)$ is the joint probability of 
$\Pr(T_\mu \ge t)$ and that no transition event takes place during time $t$ and $t+\delta t$. To the first order in $\delta t$, the latter is given by $1-\delta t  \tr [\Gamma_{\mu}^I \rho^I_{\mu \mu} +\rho^I_{\mu \mu} (\Gamma_{\mu}^I)^\dagger]$. Taking the limit $\delta t \to 0$, one then deduces $\partial_t \Pr(T_\mu \ge t) = - \Pr(T_\mu \ge t) \tr [\Gamma_{\mu}^I \rho^I_{\mu \mu} +\rho^I_{\mu \mu} (\Gamma_{\mu}^I)^\dagger]$, which is formally solved by 
\begin{equation}
    \Pr(T_\mu \ge t) = \exp \left( -\int_{0}^{t} \d t' \tr [\Gamma_{\mu}^I \rho^I_{\mu \mu} +\rho^I_{\mu \mu} (\Gamma_{\mu}^I)^\dagger]
    \right).
    \label{eq:persistent-time}
\end{equation}}

However, in order to carry out this integration, one needs the evolution of the block $\rho^I_{\mu \mu}$ of the quantum system $\mathrm{Q}$ over time, conditioned on the fact that system $\mathrm{D}$ remains in state $\ket{\mu}$. Since the ultradecohered system $\mathrm{D}$ is in state $\ket{\mu}$ until time $t$, one has $p_{\mu}=1$, and therefore the density operator $\rho^I_{\mu \mu}$ of system $\mathrm{Q}$ for this duration must be such that $\tr(\rho^I_{\mu \mu})=1$. It follows also that $\rho^I_{\nu \nu} = 0$ for all $\nu \ne \mu$. The evolution equation~\eqref{eq:Q-master-full-interaction-element} then reduces to only the `loss term.' Such an equation with only the loss term can however be only valid for $\rho_{\mu \mu}^I$ strictly at $t=0$, as it explicitly decreases  $\tr(\rho^I_{\mu \mu})$. Conditioning on state $\ket{\mu}$ of the ultradecohered system $\mathrm{D}$, one must then \emph{renormalise} the state $\rho^I_{\mu \mu}$ of the quantum system $\mathrm{Q}$ to have unit trace. 
Taking all these constraints into account, the conditional state is suggested to be $\rho_{\mu \mu}^I=  \hat{\rho}^I_{\mu \mu }/{\tr(\hat{\rho}^I_{\mu \mu})}$, where $\hat{\rho}^I_{\mu \mu }$ follows the trace-decreasing dynamics
\begin{equation}
\frac{\mathrm{d} \hat{\rho}^I_{\mu \mu}}{ \mathrm{d} t} = - (\Gamma_{\mu}^I \hat{\rho}^I_{\mu \mu} +\hat{\rho}^I_{\mu \mu} (\Gamma_{\mu}^I)^\dagger).
\label{eq:back-reaction-unnormalised}
\end{equation}
This evolution $\rho_{\mu \mu}^I=  \hat{\rho}^I_{\mu \mu }/{\tr(\hat{\rho}^I_{\mu \mu})}$ reflects the back-reaction of the ultradecohered system $\mathrm{D}$ in the conditioning state $\ket{\mu}$ onto the conditional evolution of the quantum mechanical system $\mathrm{Q}$. We refer to eq.~\eqref{eq:back-reaction-unnormalised} as the back-reaction equation, and $\Gamma^I_{\mu}$ as the back-reaction operator. 
\new{Notice that this is again an assumption, which, despite being natural, can only be eventually confirmed by means of experiments.}

\new{
Using the back reaction equation~\eqref{eq:back-reaction-unnormalised}, eq.~\eqref{eq:persistent-time} can be written as $\Pr(T_\mu \ge t) = \exp \{ \int_{0}^{t} \mathrm{d} t'  [\partial_{t'} \tr (\hat{\rho}^I_{\mu \mu})] / \tr (\hat{\rho}^I_{\mu \mu})\}$, which can be integrated to}
\begin{equation}
    \Pr(T_{\mu} \ge t) = \tr (\hat{\rho}_{\mu \mu}^I).
    \label{eq:persistent-time-final}
\end{equation}
Remarkably, this formula for the persistent time is in full similarity to that obtained by unravealling an assumed open dynamics of a system that is measured in the farfield emission~\cite{landi_current_2024}; here, it is shown that~eq.~\eqref{eq:persistent-time-final} holds for an \emph{in situ} model of measurement devices.

\new{
As for the first step distribution, one notices that the probability that the transition from the initial state $\ket{\mu}$ to state $\ket{\nu}$ takes place during time $t$ and $t + \delta t$ before any other transitions to the first order of $\delta t$  can be computed as $\delta t \tr (F^I_{\nu \mu} \rho^I_{\mu \mu}) \Pr(T_{\mu} \ge t)$.
The first step distribution is then obtained by integrating this probability over time
$\pi^{(1)}_{\nu \mu} = \int_0^{+\infty} \d t \tr (F^I_{\nu \mu} \rho^I_{\mu \mu}) \Pr(T_{\mu} \ge t)$. Using eq.~\eqref{eq:persistent-time-final} for $\Pr(T_{\mu} \ge t)$ and $\rho^I_{\mu \mu}= \hat{\rho}^I_{\mu \mu}/\tr (\hat{\rho}^I_{\mu \mu}) $, one obtains}
\begin{equation}
    \pi^{(1)}_{\nu \mu} = \int_0^{+\infty} \d t \tr (F_{\nu \mu}^I \hat{\rho}_{\mu \mu}^I).
    \label{eq:first-step-distribution}
\end{equation}

Another interesting question then arises: what is the state of system $\mathrm{Q}$ right after system $\mathrm{D}$ underwent a transition from state $\ket{\mu}$ to state $\ket{\nu}$? 
According to eq.~\eqref{eq:Q-master-classical}, if the density operator at time $t$ has only component $\rho^I_{\mu \mu}$, the density operator $\rho^I_{\nu \nu}$ at $t+\delta t$ with infinitesimal $\delta t$ can be considered to be infinitesimally small, namely
    $\rho_{\nu \nu}^I \propto  (V^{I}_{\nu \mu} \rho^I_{\mu \mu} K^I_{\mu \nu} + K^I_{\nu \mu} \rho^I_{\mu \mu} V^I_{\mu \nu})   \delta t$.
The post-transition state is suggested to be the renormalisation of this density operator,
\begin{equation}
    \rho_{\nu \nu}^I = \frac{V^{I}_{\nu \mu} \rho^I_{\mu \mu} K^I_{\mu \nu} + K^I_{\nu \mu} \rho^I_{\mu \mu} V^I_{\mu \nu}}{\tr(V^{I}_{\nu \mu} \rho^I_{\mu \mu} K^I_{\mu \nu} + K^I_{\nu \mu} \rho^I_{\mu \mu} V^I_{\mu \nu})}.
    \label{eq:post-measurement-state-general}
\end{equation}

\section{Model for von Neumann ideal measurements} 
Following von Neumann, it is assumed that the measurement device has `ready' state $\ket{0}$ and other pointer states $\ket{\mu}$ for $\mu \in \{1,2,\ldots,M\}$. They are assumed to have the same decoherence rates $\gamma_{\mu} = \gamma$, which is sufficiently large so that the measurement device is in the ultradecoherence limit.

During the measurement, the device (denoted as $\mathrm{D}$) is coupled to a quantum system to be measured (denoted as $\mathrm{Q}$), in the same way as considered above. For simplicity, it is assumed that the free dynamics of the measurement device as well as the measured system can be ignored during the measurement process, $H_{\mathrm{D}}=H_{\mathrm{Q}}=0$; there is thus no difference between the interaction picture and Schr\"odinger's picture. For each of the pointer states $\ket{\mu}$ ($\mu \ge 1$), $\ket{s_\mu}$ denotes the corresponding state of the measured system $\mathrm{Q}$ that it probes, which form a complete basis. As mentioned, one is only concerned with the first measurement clicks. 

The coupling term of the measurement device $\mathrm{D}$ to the measured system $\mathrm{Q}$ is modelled by
\begin{equation}
    V = \sum_{\mu=1}^{M} g (\ketbra{\mu}{0} + \ketbra{0}{\mu}) \otimes \ketbra{s_\mu}{s_\mu},
    \label{eq:von-neumann-coupling}
\end{equation}
where $g$ is the coupling constant, uniform for all pointer states $\ket{\mu}$ ($\mu \ge 1$).
In the notation of the previous section, $V_{\mu 0} = g \ketbra{s_\mu}{s_\mu} = V_{0\mu }$.
In this case, one directly obtains the resonant formula for $K_{\mu 0}$,
    $K_{\mu 0} =g/\gamma \ketbra{s_\mu}{s_\mu} $,
and therefore 
    $F_{\mu 0} = 2 \chi \ketbra{s_\mu}{s_\mu}$, where $\chi = { g^2}/{\gamma}$. 
Noticing that $\{\ket{s_\mu}\}_{\mu=1}^{M}$ form a complete basis for the measured system, the back-reaction operator can be found to be proportional to the identity operator,
$\Gamma_0 = \chi  \openone$.

Suppose one subjects the quantum mechanical system $\mathrm{Q}$ in state $\rho_{00}$ to the measurement device prepared in the ready state $\ket{0}$. According to the back-reaction eq.~\eqref{eq:back-reaction-unnormalised} with $\Gamma_{0}= \chi \openone$, the renormalised conditional state $\rho_{00}$ remains then constant as long as the measurement device stays in the ready state $\ket{0}$. 
It follows also, that the transition rates to a pointer state $\ket{\mu}$, $W_{\mu 0}= 2 \new{\chi} \tr(\rho_{00} \ketbra{s_\mu}{s_\mu})$, are then all time-independent. 
Also, right after the device clicks at $\ket{\mu}$, according to eq.~\eqref{eq:post-measurement-state-general}, the state of the system assumes
$\ketbra{s_{\mu}}{s_{\mu}}$. 
In other words, the measurement device follows the dynamical Born rules~\eqref{eq:continuous-measurement-rate} and~\eqref{eq:continuous-measurement-state} with $R_{\mu 0} = \sqrt{2 \chi} \ketbra{s_\mu}{s_\mu}$ as announced.

Further, the tail distribution for the persistent time of the measurement device at the ready state, $T_0$, can be easily computed using eq.~\eqref{eq:persistent-time}, yielding  $\Pr(T_0 \ge t) = \exp( - 2 \chi t)$. 
The probability for the measurement device to eventually clicks at outcome $\ket{\mu}$ in the first click can be then explicitly computed using eq.~\eqref{eq:first-step-distribution}, yielding
    $\pi_{ \mu 0}^{(1)} = \tr (\rho_{00} \ketbra{s_\mu}{s_\mu})$, in accordance with the static Born rule for the probabilities of measurement clicks. 

\section{Model for standard single photon detectors}
Widely used in modern physics and technology, single photon detectors are perhaps the most popular measurement devices in practice. 
Curiously enough, photon detectors as measurement devices are treated rather differently from the von Neumann ideal measurements~\cite{glauber_quantum_1963}.

\new{In textbooks of quantum optics, a photon detector is typically modelled as a two-state system, `0' for `ready' and `1' for `clicked'~\cite{glauber_quantum_1963,gerry_introductory_2004}. This two-state system is coupled to the quantum electromagnetic field by means of the dipole coupling. The Fermi Golden Rule is then invoked to compute the transition rate of the detector from `ready' to `clicked.' Although this approach allows for a qualitative explanation of the fact that the detector clicks at a rate proportional to the intensity of the light~\cite{glauber_quantum_1963,gerry_introductory_2004}, it is not satisfying.
Indeed, in computing the transition rate using the Fermi Golden Rule, one in fact already pre-assumes that the device behaves as a classical stochastic system without giving a particular reason.}

Here, it is suggested that photon detectors can also be modelled as measurement devices in the ultradecoherence limit, in full similarity to von Neumann ideal measurements.
Specifically, a photon detector can be modelled as a quantum system in the ultradecoherence limit with two states $\ket{0}$ (`ready') and $\ket{1}$ (`clicked').
For simplicity, it is assumed that the detector is narrow-banded and polarised, i.e., it couples to  electromagnetic field modes $r$ in a particular polarisation and in a small frequency window.
The coupling term written directly in the interaction picture is, as suggested in textbooks~\cite{gerry_introductory_2004},
\begin{equation}
    V^I(t) =  e^{i \Omega_{10} t} \ketbra{1}{0} \otimes E^+(t)   + 
       e^{i \Omega_{01} t}
     \ketbra{0}{1} \otimes E^{-}(t)  ,
    \label{eq:photo-detector-interaction-term}
\end{equation}
where $E^+(t)$ and $E^{-}(t)$ are the positive and the negative frequency components of the electromagnetic field operator. 
In the narrow-band approximation, one has~\cite{loudon_quantum_2000}
$E^+(t) = \sum_{r} g_r e^{-i \omega_r t} a_r$ and $E^{-}(t) = \sum_{r} g_r e^{+i \omega_r t} a_r^\dagger $, 
where $a_r$ and $a^\dagger_r$ are annihilation and creation operator of mode $r$,  and  all relevant constant factors were absorbed into the narrow-band mode-dependent coupling constants $g_r$.
Identifying the device transition coupling operators as $V_{10}^I = E^+(t)$  and
   $V_{01}^I = E^-(t)$, one finds
\begin{equation}
    K_{10}^I = \sum_{r}  g_r \frac{e^{-i \omega_r t} a_r}{\gamma_{10} + i (\Omega_{10} - \omega_r)},
\end{equation}
and $K_{10}^I = (K_{01}^I)^{\dagger}$. Recall that $\gamma_{10}$ denotes the average decoherence rate of the two states $\ket{1}$ and $\ket{0}$ of the detectors and $\Omega_{10}$ denotes their energy difference.
Since only modes within a narrow band (in comparison to $\gamma_{01}$) are coupled to the device, i.e., $g_r$ is sharply peaked around $\omega_r \approx \Omega_{01}$ \new{(resonant assumption)}. Within this range, one can assume that the slowly varying function $1/[\gamma_{10} + i (\Omega_{10} - \omega_r)] \approx 1/\gamma_{10}$ remains constant. 
One then finds the resonant formula
    $K_{10}^I \approx {E^+(t)}/{\gamma_{10}}$ and $
    K_{01}^I \approx {E^-(t)}/{\gamma_{01}}$.
It is readily seen  that the clicking rate of the detector is proportional to the expectation value of the intensity operator of the electromagnetic field, 
$W_{10} =   2 / \gamma_{01} \tr[ \rho_{00}^I E^-(t) E^+(t)]$, as expected. 
After the detector clicks, the state of the light is $E^+(t) \rho_{00}^I  E^-(t)/\tr[ E^+(t) \rho_{00}^I  E^-(t)]$. The photon detector thus also follows the dynamical Born rules~\eqref{eq:continuous-measurement-rate} and~\eqref{eq:continuous-measurement-state} with $R_{\mu 0}^I= \sqrt{2/ \gamma_{01}} E^+(t)$. 
These rules are of course familiar to quantum optics~\cite{glauber_quantum_1963,gerry_introductory_2004,loudon_quantum_2000}; here, it is shown that they can be derived within the ultradecoherence model of measurement devices.

\section{Time distribution of detector clicks}
Being essentially classical, ultradecohered measurement devices also allow for the computation of the time distribution of the first measurement clicks. 
This is illustrated in a simplified model of the arrival of a particle at an ultradecohred detector~\cite{muga_time_2008,muga_time_2009}. 
To this end, one considers a simplified model of a tight binding system on two sites $\ket{\mathrm{L}}$ (left) and $\ket{\mathrm{R}}$ (right), as also sometimes investigated in the standard literature~\cite{muga_time_2008,muga_time_2009}. 
A particle is initiated at site $\ket{\mathrm{L}}$, which can tunnel between the two sites $\ket{\mathrm{L}}$ and $\ket{\mathrm{R}}$ with kinetic energy $-\Delta$. 
Effectively, one obtains a two-state system with Hamiltonian
$H= -{\Delta} \sigma_x$ in the basis $\{\ket{\mathrm{L}},\ket{\mathrm{R}}\}$,
where $\sigma_x$ denotes the $x$-Pauli matrix.
At site $\ket{\mathrm{R}}$, a detector is placed to probe the particle. 
The coupling to the detector transition from state `ready' ($\ket{0}$) to `clicked' ($\ket{1}$) is given by
$V_{10} = g \ketbra{\mathrm{R}}{\mathrm{R}}$, where $g$ is the coupling constant.
With this model, the back-reaction operator is obtained in Schr\"odinger's picture as 
$\Gamma_{0} = \chi \ketbra{\mathrm{R}}{\mathrm{R}}$,
where $\chi=g^2/\gamma$, with $\gamma$ being the average decoherence rate of the states `ready' $\ket{0}$ and `clicked' $\ket{1}$ of the detector. 
The back-reaction evolution can be written in Schr\"odinger's picture as 
\begin{equation}
    \frac{\mathrm{d} \hat{\rho}_{00}} {\mathrm{d} t} = -i (H_{\mathrm{eff}} \hat{\rho}_{00} -  \hat{\rho}_{00} H_{\mathrm{eff}}^\dagger)\new{,}
\end{equation}
with the non-hermitian Hamiltonian
$H_{\mathrm{eff}} = -{\Delta} \sigma_x -  i \chi \ketbra{\mathrm{R}}{\mathrm{R}}$. 
This non-hermitian evolution is familiar to the literature since Allcock's seminal work~\cite{allcock_time_1969-1}; see also~\cite{muga_time_2008,muga_time_2009} and the references therein.
As well-known, the solution to the unnormalised back-reaction equation~\eqref{eq:back-reaction-unnormalised} can be written as $\hat{\rho}_{00}= \ketbra{\hat{\psi}(t)}{\hat{\psi}(t)}$, where $\ket{\hat{\psi}(t)} = e^{- i H_{\mathrm{eff}} t } \ket{\mathrm{L}}$.
Using eq.~\eqref{eq:persistent-time}, one can then explicitly find the tail distribution of time of arrival as $\Pr(T_0 \ge t) = e^{- \chi t} \left({4\Delta^2}  - {\chi^2} \cos \omega t + \chi \omega \sin \omega t \right)/\omega^2$ where $\omega = \sqrt{4 \Delta^2 - \chi^2}$. While the resulted distribution is consistent with the standard literature~\cite[page 137]{muga_time_2009}, it is to be emphasized that its parameter $\chi$ is here explicitly expressed in terms of the coupling constant $g$ and the decoherence rate $\gamma$ characterising the ultradecohered detector.

\section{Perspective for the foundations of quantum mechanics}
\label{sec:foundations}
\new{
Equation~\eqref{eq:von-neumann-sum} is often the starting point of many debates around the nature of quantum measurements, expanding also to various related foundational issues of quantum mechanics, {such as the interpretation of Schr\"odinger's cat states~\cite{schrodinger_gegenwartige_1935}, the existence of many worlds~\cite{everett_relative_1957}, or the implication of Wigner's friend~\cite{wigner_remarks_1995,deutsch_quantum_1985}}; see also ~\cite{schlosshauer_decoherence_2007,schlosshauer_decoherence_2005,schlosshauer_elegance_2011,zurek_decoherence_2003,zeh_interpretation_1970,zeh_toward_1973,zurek_pointer_1981,zurek_environment-induced_1982,kiefer_quantum_2022,presilla_measurement_1996}.
Among these is the question of finding an argument for reducing from the premeasurement state~\eqref{eq:von-neumann-sum} to a particular term in the decomposition, say,  $\ket{\mu} \otimes \ket{s_\mu}$, which eventually describes the perception of a definite outcome by the observer---\emph{the problem of definite outcomes}~\cite{schlosshauer_decoherence_2005,zurek_decoherence_2003}. 
The problem appears even more severe once it was realised that a basis transformation of the measurement device results in another decomposition with the same appearance as eq.~\eqref{eq:von-neumann-sum}~\cite{zurek_pointer_1981}. 
It is thus necessary first to point out why the decomposition of the joint state into the pointer states of the measurement devices by eq.~\eqref{eq:von-neumann-sum} is preferred---\emph{the preferred basis problem}~\cite{schlosshauer_decoherence_2005,zurek_decoherence_2003}.

A breakthrough in understanding these problems has been made in the seventies~\cite{zeh_interpretation_1970,zeh_toward_1973,zurek_pointer_1981,zurek_environment-induced_1982}, marking the initiation of the so-called decoherence program~\cite{schlosshauer_decoherence_2005,zurek_decoherence_2003,kiefer_quantum_2022}. 
It was realised that the environment surrounding the measurement device plays a crucial role in defining its behaviour. 
In particular, due to the specific way it couples to the environment, most states of the measurement device are unstable, as they quickly get entangled with the environment, thus `decohered'~\cite{zeh_interpretation_1970,zeh_toward_1973,zurek_pointer_1981,zurek_environment-induced_1982}. 
Exceptions are those that are eigenstates of the coupling to the environment (or are approximately so), which are then selectively stable under the interaction~\cite{zurek_pointer_1981,zurek_pointer_1981,zurek_environment-induced_1982}. 
These states form the preferred basis for the measurement device~\cite{zurek_pointer_1981,zurek_environment-induced_1982,zurek_decoherence_1991,zurek_decoherence_2003}.
The main recognised contribution of the decoherence program is an explanation of the preferred basis~\cite{schlosshauer_decoherence_2005,schlosshauer_quantum_2019}. 
To which extent it clarifies the problem of definite outcomes remains controversial~\cite{schlosshauer_decoherence_2005,schlosshauer_decoherence_2007,schlosshauer_quantum_2019}; see also more recent developments~\cite{zurek_quantum_2018,zurek_quantum_2022,kiefer_quantum_2022}.

The ultradecoherence model of measurement devices assumes a preferred basis imposed by a fast dephasing dynamics. It seems, however, that the premeasurement state~\eqref{eq:von-neumann-sum} does not necessarily arise at the local level between the measured system and the measurement device. \new{Instead, the measurement device always assumes a definite state upto the accuracy set by the fast decoherence timescale.}
\new{Notice that even when one would like to extend to consider the global unitary of the measured system, the measurement device \emph{and} the environment, it is not clear how the premeasurement state~\eqref{eq:von-neumann-sum} can be justified.}
\new{This hints at a new interesting perspective for the foundations of quantum mechanics}.
}

\hide{
Although artificial, one may want to hypothetically isolate a large environment together with the measurement device and consider the unitary evolution of this `super measurement device' and the measured system. However, even in this case, one cannot assume that~\eqref{eq:von-neumann-sum} holds without a careful investigation. Indeed, as we have seen for the measurement device itself, the unitary dynamics of a system after isolation from its environment cannot be easily extrapolated from its behaviour in the presence of the environment without further justification. 


Measurement devices are not the only systems who are in ultradecoherence. Schr\"odinger's cat~\cite{schrodinger_gegenwartige_1935}, as a macroscopic system, can also be expected to be in the ultradecoherence limit. It remains so even in an isolated room, as long as the accompanying environment is large enough for its ultradecoherence limit to establish. As such, it will undergo a classical transition from `alive' to `death' at a random time by the perturbation induced by a decaying quantum particle in the room. Accordingly, the particle gets a transition into the definite state of `decayed.' Such a picture is so natural that one might have expected it from the outset. 
The price to pay for such a natural picture is that the cat itself, or more generally, our reality, is only classically defined upto a certain timescale. It is however also reasonable to believe that trying to interfere with how the cat interacts with its environment at a timescale shorter than its decoherence limit would interrupt its living functions to the point that it might not be `death' nor `alive.' This effective reality upto a certain timescale imposed by the decoherence caused by the surrounding environment is implied in our approach. On the one hand, this effective picture of reality has been present since the inception of the decoherence theory. On the other hand, it seems to carry certain disappointment that its reception might remain difficult.

Again, one may still insist to extend to consider an `absolute description' of the cat and the environment as a whole in a single pure state. This, as also for measurement devices, appears to be rather artificial. On the one hand, the local behaviour of the cat itself will not change, as long as its environment is large enough for its ultradecoherence to establish. On the other hand, such a global description must take into account the detailed new dynamics of the system under the interference of the isolation, and the state of the from~\eqref{eq:von-neumann-sum} may not be assumed without a careful justification; see the discussion on the problem of definite outcomes in the main text.
}
\section{Conclusion}
This investigation gives a glimpse of the hidden intriguing physics of the measurement processes in quantum mechanics. 
The ultradecoherence model is believed to be just one but many ways to study them.
On the one hand, investigation of the ultradecoherence phenomenon for various more realistic measurement setups presents an interesting direction for future research. 
The implication of the model for the foundations of quantum mechanics is an interesting subject for discussion.
Comparison of the \emph{in situ} prediction on the time distribution of the first detector clicks with those obtained from the farfield theory~\cite{landi_current_2024} and other methods~\cite{muga_time_2008,muga_time_2009} might also contribute to the understanding of the time problem in quantum mechanics.
On the other the hand, exploring the dynamics beyond the assumption of ultradecoherence holds a promise for better understanding of the quantum-to-classical transition at measurement devices.

\bmhead{Acknowledgements}
The author thanks
Son Cao,  
Assegid M. Flatae,
Claus Kiefer,
Matthias Kleinmann,
Roberto Onofrio,
Leonardo S. V. Santos,
Maximilian Schlosshauer,
Konrad Szyma\'nski,
and 
Wojciech H. Zurek
for helpful discussions.
The clarity of this manuscript greatly benefited from the comments from 
Fabian Bernards, 
Otfried G\"uhne, 
V. Lien Nguyen,
and
Stefan Nimmrichter.
The International Center for Interdisciplinary Science and Education (ICISE, Quy Nhon) is acknowledged for supporting the author's visit at the Neutrino Group in Quy Nhon, during which he had chance to see a photon detector.



\bmhead{Funding}
This work is supported by 
the Deutsche Forschungsgemeinschaft (DFG, German Research Foundation, project numbers 447948357 and 440958198), 
the Sino-German Center for Research Promotion (Project M-0294), 
the German Ministry of Education and Research (Project QuKuK, BMBF Grant No. 16KIS1618K),
and the EIN Quantum NRW.



\bibliography{ultra-decoherence}

\end{document}